\documentclass{amsart}

\usepackage{fullpage,amssymb,epsf}

\newcommand{\be}{\begin{equation}}
\newcommand{\ee}{\end{equation}}
\newcommand{\bea}{\begin{eqnarray}}
\newcommand{\eea}{\end{eqnarray}}
\newcommand{\beas}{\begin{eqnarray*}}
\newcommand{\eeas}{\end{eqnarray*}}

\def\photon{\;\raisebox{0mm}{\epsfxsize=5mm\epsfbox{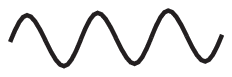}}\;}
\def\fermion{\;\raisebox{0mm}{\epsfxsize=5mm\epsfbox{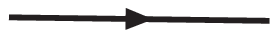}}\;}
\def\vertex{\;\raisebox{-1mm}{\epsfxsize=6mm\epsfbox{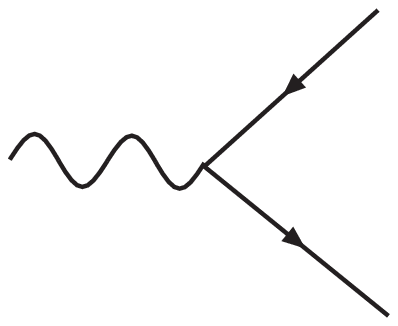}}\;}
\def\photonk{\;\raisebox{-7mm}{\epsfysize=16mm\epsfbox{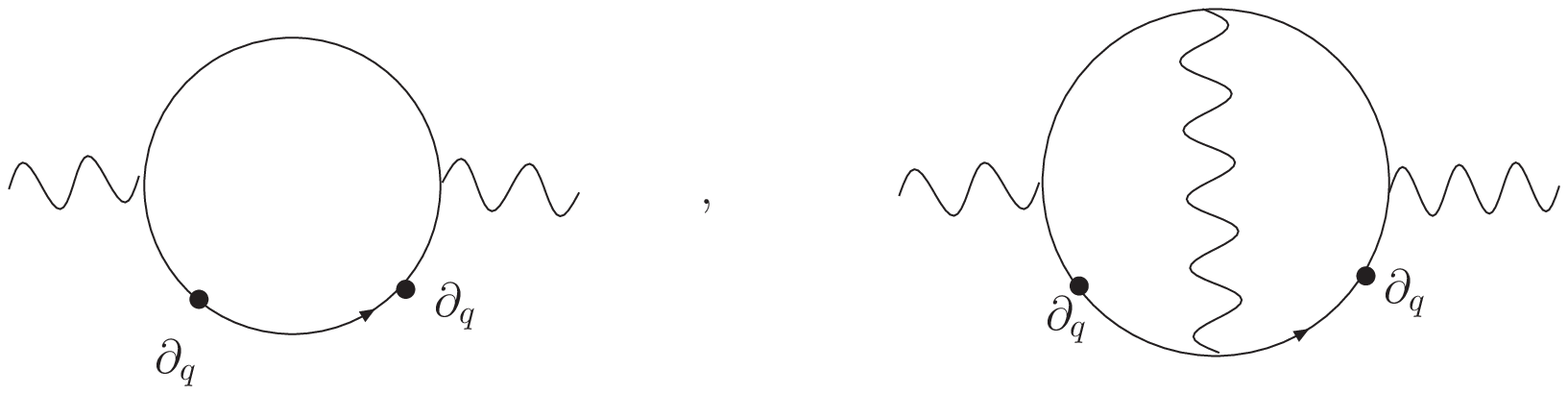}}\;}
\def\fermionk{\;\raisebox{-7mm}{\epsfysize=16mm\epsfbox{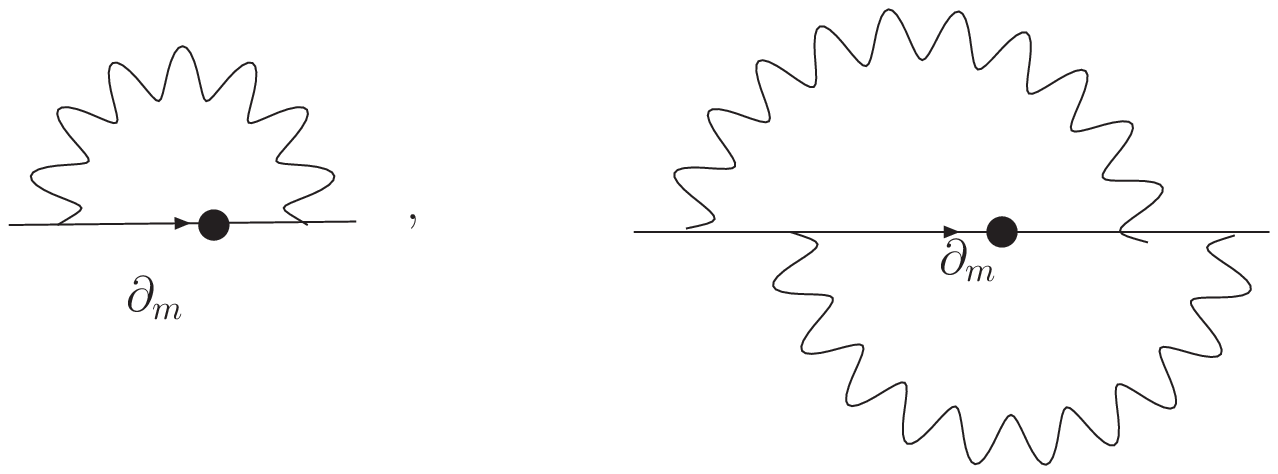}}\;}
\def\vertexk{\;\raisebox{-7mm}{\epsfysize=16mm\epsfbox{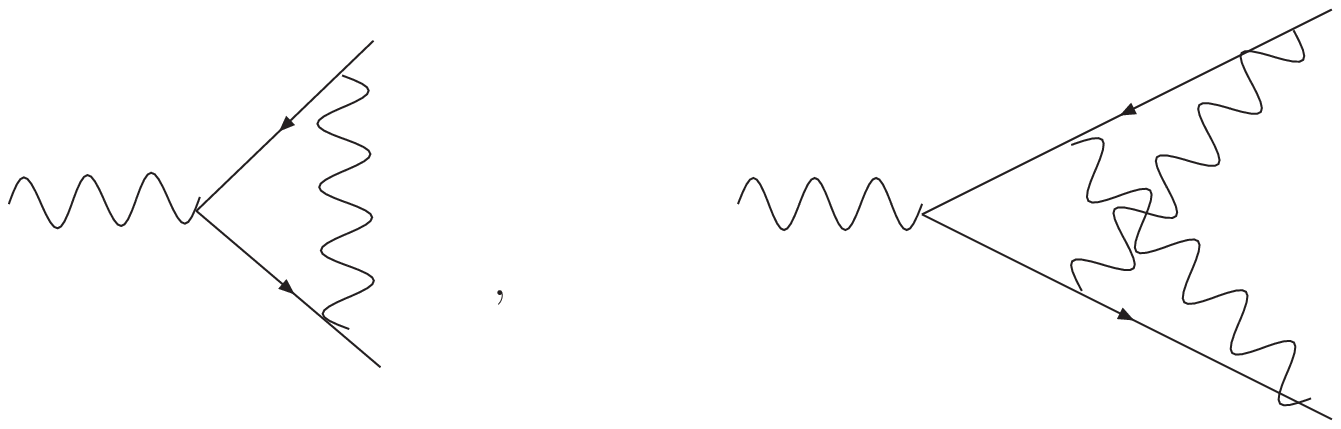}}\;}
\def\vertexks{\;\raisebox{-15mm}{\epsfysize=40mm\epsfbox{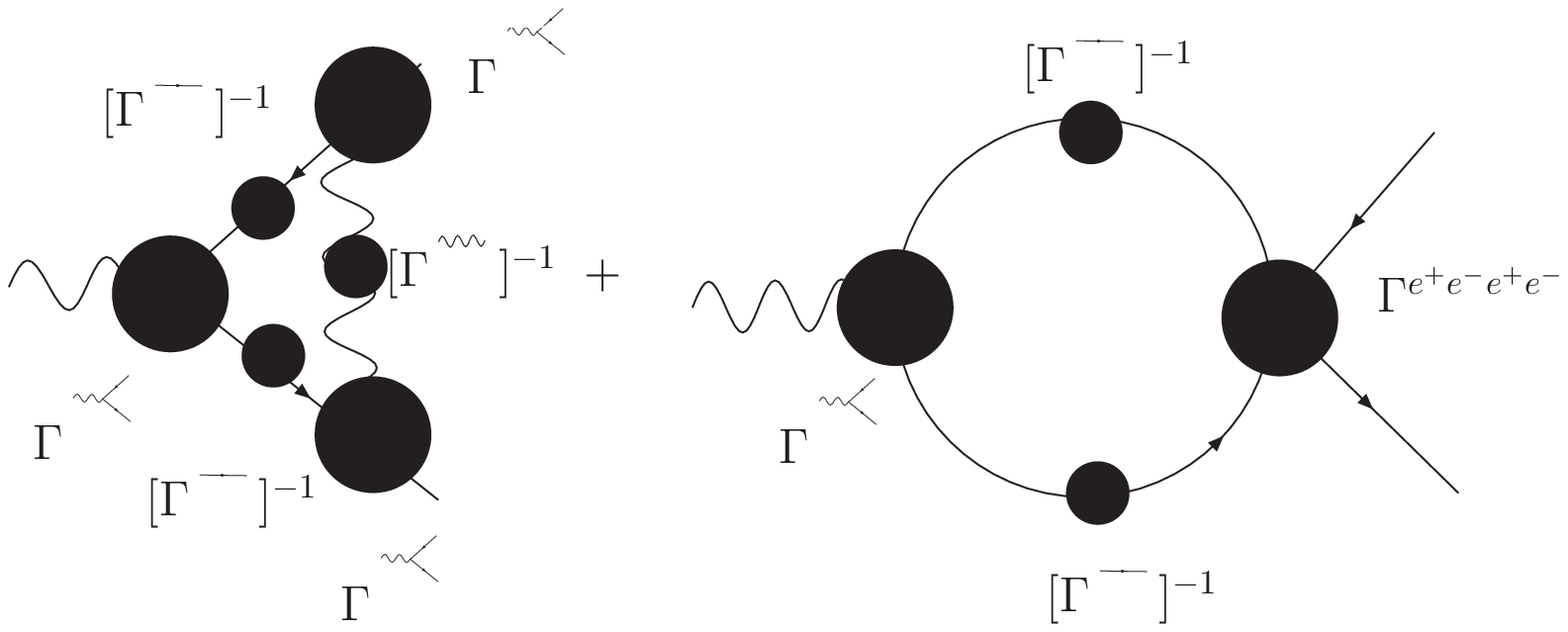}}\;}

\def\qeds{\;\raisebox{-4mm}{\epsfysize=12mm\epsfbox{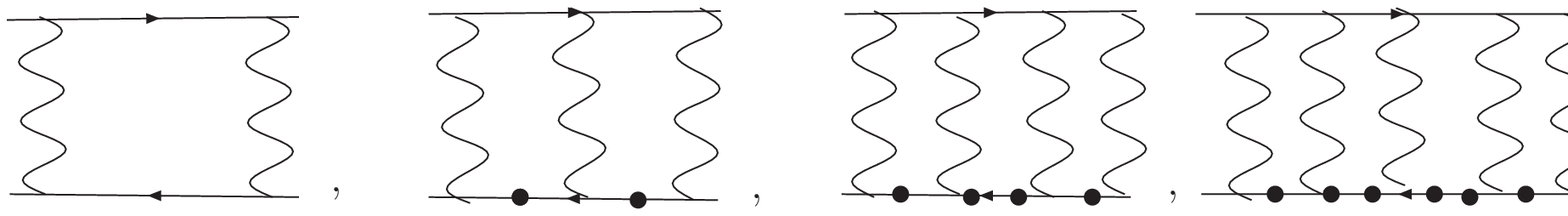}}\;}
\def\gravcarry{\;\raisebox{-8mm}{\epsfysize=24mm\epsfbox{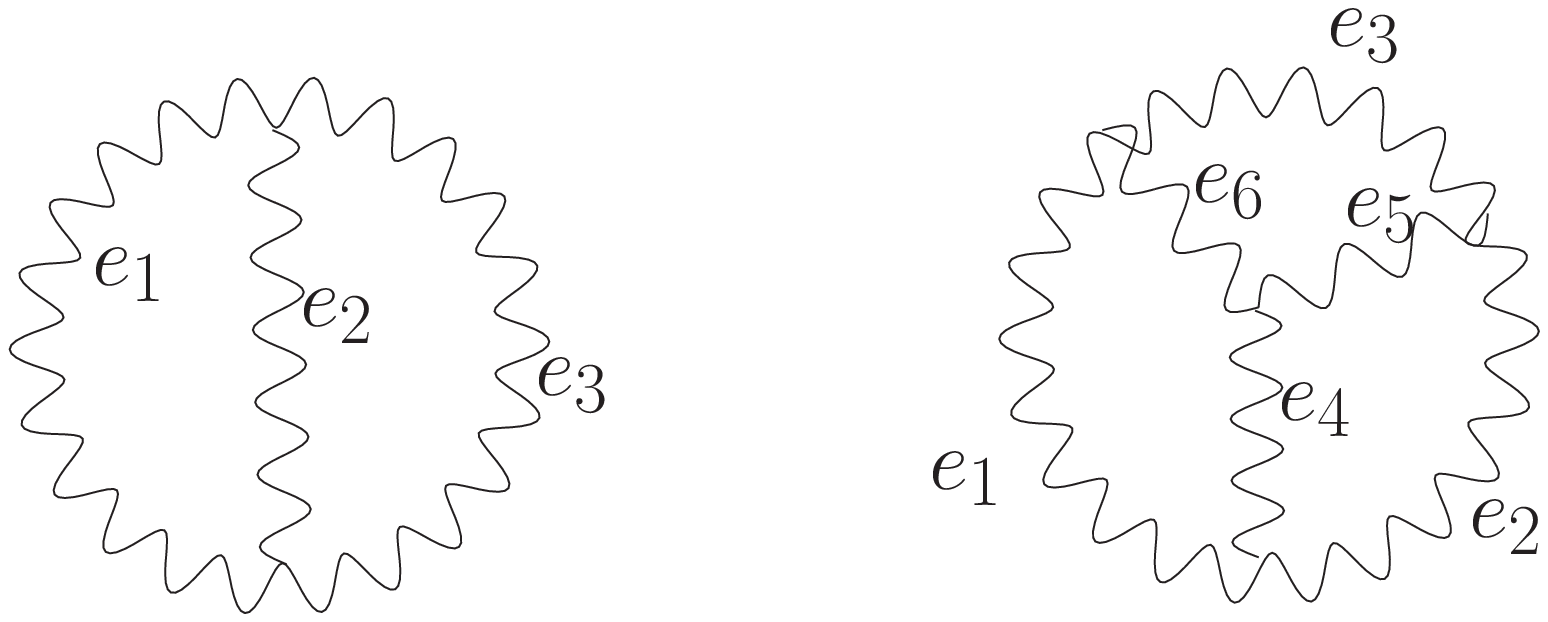}}\;}
\def\Qins{\;\raisebox{-16mm}{\epsfysize=48mm\epsfbox{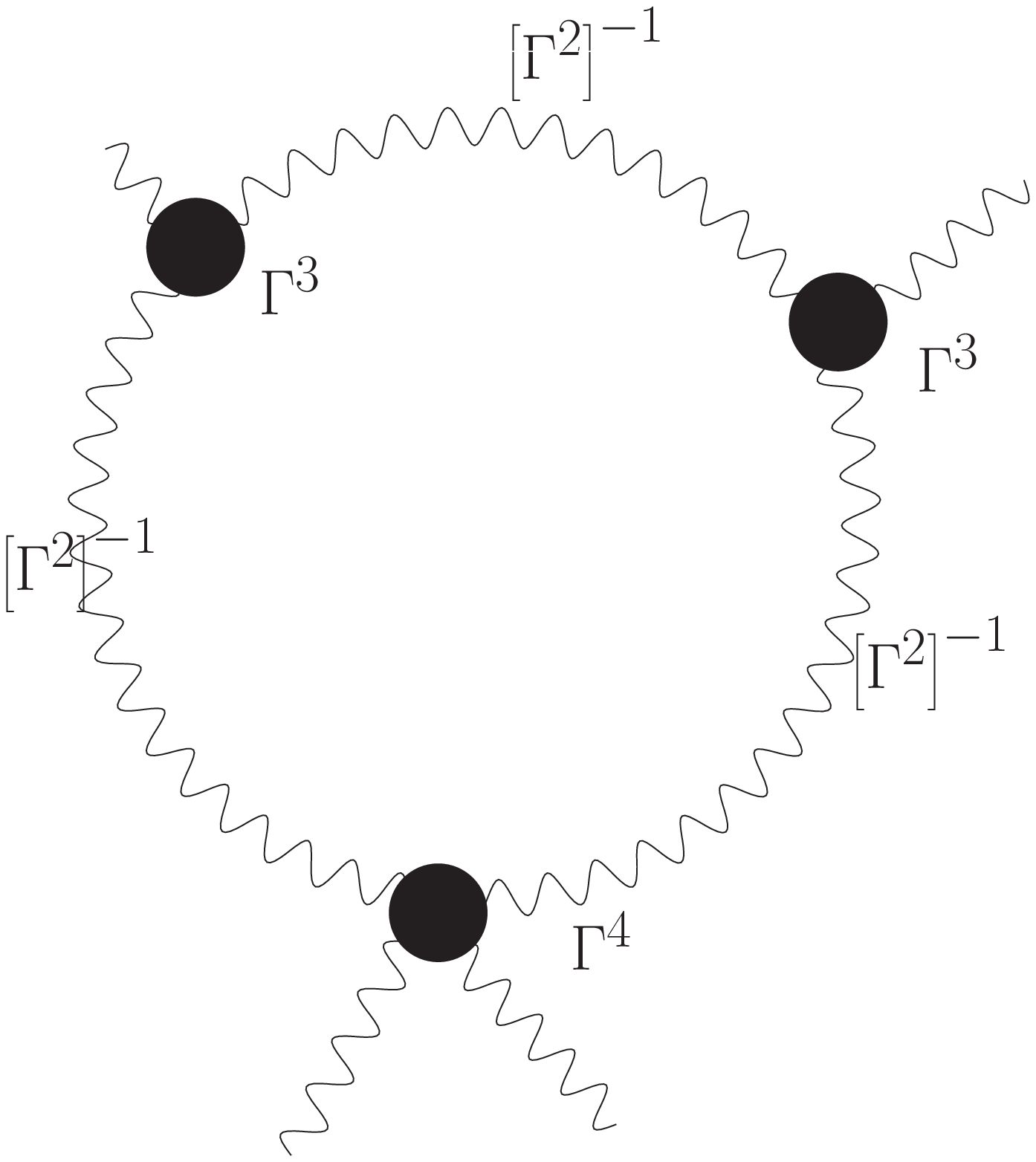}}\;}

\newtheorem{thm}{Theorem}

\newtheorem{prop}[thm]{Proposition}
\newtheorem{cor}[thm]{Corollary}

\title{A remark on quantum gravity${}^*$}
\author{Dirk Kreimer}
\address{kreimer@ihes.fr, IHES (http://www.ihes.fr) and Boston U.\ (http://math.bu.edu)}
\thanks{${}^*$Talk given at "NCG-Conference in honor of Alain Connes", Paris, 29/03/07-06/04/07, dedicated to Alain in friendship and gratitude.
Work supported in parts by grant NSF-DMS/0603781. Author supported by CNRS}
\begin{document}
\maketitle
\begin{abstract}
We discuss the structure of Dyson--Schwinger equations in quantum
gravity and conclude in particular that all relevant skeletons are
of first order in the loop number. There is an accompanying sub Hopf
algebra on gravity amplitudes equivalent to identities between
n-graviton scattering amplitudes which generalize the Slavnov Taylor
identities. These identities map the infinite number of charges and
finite numbers of skeletons in gravity to an infinite number of
skeletons and a finite number of charges needing renormalization.
Our analysis suggests that gravity, regarded as a probability
conserving but perturbatively non-renormalizable theory,  is
renormalizable after all, thanks to the structure of its
Dyson--Schwinger equations.
\end{abstract}
\section{Introduction}
A renormalizable theory poses a computational problem for a
theoretical physicist: even if only a finite number of amplitudes
need renormalization, the quantum equations of motion -the
Dyson--Schwinger equations (DSE)- ensure that these amplitudes must
be calculated as iterated integrals based on a skeleton expansion
for the Green functions. There is an infinite series of skeletons,
of growing computational complexity, and thus a formidable challenge
at hand. Order is brought to this situation by the fact that the
skeletons can be organized in terms of the underlying Hochschild
cohomology of the Hopf algebra of a renormalizable theory, the
computational challenge remains though in the analytic determination
of the skeletons and their Mellin transforms \cite{tor,KY1,KY2}.
This approach, combining the analysis of the renormalization group
provided in \cite{RHII} with the analysis of the mathematical structure of DSE
provided in \cite{BroadhK,BergbK,tor}, has led to new methods in solving DSE beyond perturbation theory
\cite{BroadhK,KY1,KY2}.

A nice fact is that internal symmetries can be systematically
understood in terms of this Hochschild cohomology: Slavnov--Taylor
identities are equivalent to the demand that multiplicative renormalization is
compatible with the cohomology structure \cite{anatomy}, leading to
the identification of Hopf ideals generated by these very Ward and
Slavnov--Taylor identities \cite{walter}.

For a non-renormalizable theory the situation is worse: the
computational challenge for the theorist is repeated infinitely as
there is now an infinite number of amplitudes demanding
renormalization, each of them still based on an infinite number of
possible skeleton iterations.

But the interplay with Hochschild cohomology leads to surprising new
insights into this situation, which in this first paper we discuss
at an elementary level for the situation of pure gravity.
\subsection*{Acknowledgments}
It is a pleasure to thank David Broadhurst, John Gracey and Karen
Yeats for discussions.
\section{The structure of Dyson--Schwinger Equations}
To compare the situation for a renormalizable QFT with the situation for an
unrenormalizable one, we consider QED in four and six
dimensions of spacetime.
\subsection{QED${}_4$}
Let us consider as a typical example quantum electrodynamics in four
dimensions. The DSE involves a sum over superficially convergent
skeleton kernels which are in one-to-one correspondence with the
primitives of the Hopf algebra underlying perturbation theory. There
is an infinite number of primitive skeleton graphs, a finite
number of them at each loop order. Furthermore, there is a finite number of
monomials, tree-level amplitudes represented graphically as elements of a finite set $\mathcal R$
\be {\mathcal R}=\left\{\photon,{\fermion\atop m},{\fermion
\atop \partial},\vertex\right\}, \ee $|{\mathcal R}|=4<\infty$. The
Lagrangian is \be {\mathcal L}=\sum_{r\in{\mathcal
R}}\hat{\phi}(r)=-\bar{\psi}[\partial\!\!/+ieA\!\!/]\psi-m\bar{\psi}\psi-\frac{1}{4}F^2,\ee
with coordinate space Feynman rules $\hat{\phi}$,
\be \hat{\phi}(\photon)=\frac{1}{4}F^2,\;\hat{\phi}({\fermion\atop m})=m\bar{\psi}\psi,\;\hat{\phi}({\fermion\atop \partial})=\bar{\psi}\partial\!\!/\psi,
\hat{\phi}(\vertex)=\bar{\psi}A\!\!/\psi.\ee The Dyson--Schwinger equations take the corresponding form \cite{tor}
$\forall r\in{\mathcal R}$, \be G^r(\alpha,L)=1\pm\sum_{k=1}^\infty
\alpha^k \phi_R\left(B_+^{r;k}(X^rQ^k)\right),\ee with renormalized Feynman rules $\phi_R$ in momentum space
with subtractions at an Euclidean momentum $q^2=\mu^2$, $L=\ln q^2/\mu^2$, normalized to unity on tree-level terms $\in{\mathcal R}$.
We restrict to zero-momentum transfer in vertex functions such that Green functions $G^r(\alpha,L)$ are functions of the single kinematical variable
$L$ and projection onto form-factors is understood in accordance with the set ${\mathcal R}$.

The plus sign above is taken
for the vertex functions, and the minus sign for inverse
propagators. $X^r$ is a solution to the corresponding combinatorial
DSE in Hochschild cohomology, \be X^r=\mathbb{I}\pm\sum_{k=1}^\infty \alpha^k
B_+^{r;k}(X^rQ^k),\ee with $G_R(\alpha,L)=\phi_R(X^r)$ and \be
Q_R(\alpha,L)=\phi_R(Q)=\frac{1}{\sqrt{\phi_R(X^{\photon})(\alpha,L)}},\;Q=\frac{1}{\sqrt{X^{\photon}}},\ee defines the
invariant running charge $\alpha/Q_R$. We emphasize that the above are DSE for 1PI Green functions.

The primitives \be B_+^{r;k}(\mathbb{I}), \; bB_+^{r;k}=0 \Leftrightarrow
\Delta B_+^{r;k} = B_+^{r;k} \otimes \mathbb{I}+({\rm id}\otimes
B_+^{r;k})\Delta,\ee are such that upon application of the
renormalized Feynman rules, $\phi_R(B_+^{r;k})$, they give the Dyson kernels  for the
vertex $\vertex$, the quenched $\beta$ function and light-by-light
scattering amplitudes for the photon $\photon$, suitable mass derivatives
for the fermion propagator ${\fermion\atop m}$, while the kinetic part of the
fermion propagator ${\fermion\atop \partial}$ is taken care off by the Ward identity. Here
are the primitives up to two loops: \bea r=\photon:  & & \photonk,\nonumber\\
r=\fermion:  & & \fermionk,\label{2l}\\ r=\vertex:  & & \vertexk.\nonumber\eea
Labeled dots indicate derivatives with respect to external momenta $q$ or fermion masses $m$, such as to render these integral kernels logarithmically
divergent.

In general,
\be B_+^{r;k}=\sum_{|\gamma|=k,{\rm res}(\gamma)=r}B_+^\gamma,\ee
where $B_+^\gamma$ is defined via pre-Lie insertion into the primitive $\gamma$ such that $B_+^{r;k}$ is Hochschild closed \cite{anatomy,tor}.
As a result  we
get a sub Hopf algebra which is generated by a single one-cocycle $B_+^{r;k}$ in each loop-degree $k$, \cite{KY2}.

The degree of divergence of a graph $\Gamma$ with $f$ external
fermion lines and $m$ external photon lines in $D$ dimensions is \be
\omega_D(\Gamma)=\frac{3}{2}f+m-D-(D-4)(|\Gamma|-1)\Rightarrow
\omega_4(\Gamma)=\frac{3}{2}f+m-4.\ee This is independent of the
loop number for QED${}_4$, $D=4$, and is a sole function of the
number and type of external legs. $\omega_D(\Gamma)$ determines the number of derivatives needed to render
a graph logarithmically divergent.

This finishes our summary of QED$_{4}$ as a typical renormalizable theory.

\subsection{QED${}_6$} Here, $|\{{\mathcal R}\}|=\infty$,
as each amplitude is superficially divergent in sufficiently high
loop number.

Searching for primitives, there is a maximal loop order for each amplitude after
which graphs contributing to any given amplitude have sub-divergences.
 We have $\omega_6(\Gamma)=\frac{3}{2}f+m-4-2|\Gamma|$,
and this is now a function of the number and type of external legs and
the number of loops.

For example, the Dyson kernels (the $e^+e^-\to e^+e^-$ scattering
graphs 2PI in the forward channel, \cite{BjDr}) have power-counting degree \be
\omega(\Gamma)=2-2|\Gamma|,\ee and one immediately proves that no
such graph is primitive beyond one loop.

But to set up our Dyson--Schwinger equations correctly, we need to
investigate the integral kernels obtained by taking a sufficient
number of derivatives, with respect to masses or external momenta,
so as to obtain log-divergent integral kernels. This can be
graphically indicated by (labeled) dots on the graph as in
\cite{BDK}, and as we did in (\ref{2l}) above already. One immediately proves that these top-degree
contributions provide an infinite number of skeleton kernels.  Upon taking such derivatives we create actually new integral kernels free of subdivergences. There are log-divergent kernels
for each amplitude, at each loop number.
\begin{cor}For each $r\in{\mathcal R}$ and each positive integer $k$, there exists non-trivial one-cocycles $B_+^{r;\gamma}$
with $|\gamma|=k$.\end{cor} Proof: It suffices to give, for each amplitude, a single series of dotted graphs $\Gamma$ with $\omega_6(\Gamma)$ dots each which has
a primitive member at each loop order. We do so for the example $e^+e^-\to e^+e^-$ by the following figure.  \be\qeds\ee where dots represent derivatives wrt an external momentum.
For other amplitudes similar series are easily constructed.\hfill $\Box$\\[5mm]

Summarizing, the DSE take a form similar to QED${}_4$, \bea
G^r(\alpha,L) & = & 1\pm\sum_{k=1}^\infty \alpha^k
\sum_{|\gamma|=k,{\rm res}(\gamma)=r}\phi_R\left(B_+^{\gamma}\left(X_\gamma\right)\right),\\
X_\gamma & = & \prod_{v\in\gamma^{[0]},e\in\gamma^{[1]}_{\rm int}}\frac{\Gamma^v}{\Gamma^e},\eea only that the number of amplitudes
needing renormalization, $|{\mathcal R}|$, is  infinite, and for each
amplitude we have an infinite number of kernels. Note that we can not write this in terms of $B_+^{r;k}$, as we are lacking
the crucial identity
\be X_\gamma=X_{\gamma^\prime},\ee for all primitive $\gamma,\gamma^\prime$ with the same loop number, and contributing to the same amplitude \cite{anatomy}.

Indeed, note that we obtain new primitives $\gamma$ as in the case of the initial photon vertex, which now has two one-loop kernels, one involving
a new Green function $G^{e^+e^-\to e^+e^-}$ contributing to $\mathcal R$, given here with their dressings:
\be\vertexks\ee
We would hence need a relation like
\be \frac{\left[\Gamma^{\vertex}\right]^3}{[\Gamma^{\fermion}]^2\Gamma^{\photon}}=\frac{\Gamma^{\vertex}\Gamma^{e^+e^-e^+e^-}}{[\Gamma^{\fermion}]^2},\ee
and infinitely many more of them, to render the theory renormalizable by renormalization of a single charge.

It is this double infinity which renders such a theory
non-predictive: there is an infinite amount of work for a theorist (work out the kernels) and an infinite amount of work for an experimentalist:
measure an infinite number of charges to render the theory predictive.
\section{Gravity}
We consider pure gravity understood as a theory based on a graviton propagator and $n$-graviton couplings as vertices.
A fuller discussion incorporating ghosts and matter fields is referred to future work.

An immediate observation concerns powercounting in such a theory. If we
work with Feynman rules as given in \cite{Bryce}, we see that each
$n$-graviton vertex is a quadric in momenta attached to the vertex.
This has an immediate consequence.
\begin{cor} Let $|\Gamma|=k$. Then $\omega(\Gamma)=-2(|\Gamma|+1)$.\end{cor}
Proof: A 1PI one-loop graph has as many internal edges as vertices.
Their contributions to the superficial degree of divergence hence cancel.
We conclude that one-loop graphs are quadratically divergent, in accordance with the corollary.
Increasing the loop order by one in a 1PI graph introduces one more propagator than vertices, the net gain in
powercounting is hence $-2$, $+2$ from counting
propagators and vertices, and $-4$ from counting loops. \hfill $\Box$\\
We have now a situation dual to a renormalizable theory: The
superficial degree of divergence depends on the loop number, but not
on the number and type of external legs. While in a renormalizable theory like massless QED${}_4$
we have a single amplitude which needs renormalization -thanks to the Ward identity-, and primitives at each loop order, we have found a nice loop-to-leg duality: we have primitives only at first loop order, but for any number of external legs.

Investigating further,  the crucial
result of this paper is that quantum gravity is sufficiently
non-renormalizable such that its skeletons are all one-loop.
\subsection{All skeletons are one-loop}
As powercounting is determined by the loop number instead of by the
external leg structure, there is only one two-loop graph, made of
three carrying edges. The notion of a carrying edge takes account of
the fact that the powercounting of edges and vertices cancels. We
hence ignore external edges and consider a string of edges, interrupted only by vertices with couplings to external legs, as a
single carrying edge. We extend this notion to subgraphs. External edges are not explicitly given from now on, as they play no role
for powercounting.

For a graph
$\Gamma$, we let $d_{\omega(\Gamma)}$ be the set of graphs obtained
by distributing $\omega(\Gamma)$ dots over $\Gamma$, indicating the
action of derivatives wrt to external momenta or masses. Derivatives on vertices will likewise
be indicated by dots on internal propagators in accordance with the leg which is involved.

Let us look at the the following two graphs.
\be \gravcarry.\ee
For the two loop graph on the left, we can distribute six markings over its three internal edges $e_1,e_2,e_3$ to render it log-divergent.
We have to do so such that each of the one-loop subgraphs $\{(e_1,e_2)\}$, $\{(e_2,e_3)\}$, $\{(e_3,e_1)\}$ has five markings, as it has to be finite.
But there is no partition of six into three integers such that the sum of any two of those integers is greater than four.

So there is no primitive two-loop graph. The first integer which has such a partition is $8\to (3,3,2)$.
We hence have to require that at higher loop orders, every two-loop graph, which necessarily is based on three carrying edges, obtains eight dots.

Let us now look at a three loop graph $\Gamma$. We have $\omega(\Gamma)=-8$. So we have eight markings at our
disposal. We need to distribute them in a way such that each of its
two-loop subgraphs contains all of them, as we would have to render
all its two-loop subgraphs finite. This is impossible as two such
two-loop subgraphs will not share all carrying edges. In the figure,
we have six edges $e_1,\cdots,e_6$. Any five of those six edges form a two-loop subgraph on three carrying edges.
For example edges $e_1,e_2,e_3,e_4,e_5$ form a two-loop subgraph on the three carrying edges $\{(e_1,e_3),(e_4,e_5),e_2\}$.
Distributing $3+3+2$ dots in any way over these three carrying edges leaves for example the two-loop subgraphs like
$\{(e_1,e_3),(e_4,e_5),e_6\}$, containing $e_6$, divergent.

So there is no primitive three-loop kernel, and eleven or more
markings are needed to render the three-loop graphs finite. This
argument continues in a straightforward manner and hence we will not find a primitive in $d_{\omega(\Gamma)}$
for any graph $\Gamma$ beyond first loop order. We hence have proven for any graph $\Gamma$
\begin{thm} The set $d_{\omega(\Gamma)}$ contains no primitive element.\end{thm}
This is strikingly different in comparison to a generic non-renormalizable theory like QED$_{6}$
which we discussed in the previous section.

\subsection{Relations between Green functions}
The double infinity, in loops and legs, of a non-renormalizable theory has turned for
quantum gravity into a single infinity  parameterized by the
different number of external legs.
While for a renormalizable theory, the
degree of divergence is constant over the number of loops and varies
with the number of external legs, we have the dual situation for
gravity: it is constant over the number of external legs, but varies
with the loop number.

The question we ask is if there is a map which connects the two
situations of quantum gravity and a renormalizable field theory,
using the structure of their respective DSEs? Is quantum gravity,
by such a map,
non-pertubatively renormalizable whilst being perturbatively
non-renormalizable? This thought, actively pursued in the context of the Wilsonian renormalization group since several years
\cite{Martin}, indeed emerges also from our study of the DSE of quantum gravity.

We can work out how such a map should look. What we need is a map
which gives an infinite number of relations between the infinite
number of amplitudes needing renormalization such that only a finite
number of amplitudes remain indetermined, possibly on the expense of
increasing the number of primitives infinitely.

Our input is the structure of the DSE written in terms of their
Hochschild one-cocyles, combined with the request for multiplicative
renormalization. In \cite{anatomy}, in the context of a non-abelian gauge theory, it was shown how this request
leads to the Slavnov--Taylor identities for the couplings. There,
this request led, for such a gauge theory as a renormalizable theory with a a finite set
${\mathcal R}$, to a finite number of relations ensuring that there
is a unique renormalized coupling.

The request \cite{anatomy}
\be X_\gamma=X_{\gamma^\prime}\ee
leads to a Hopf ideal in the pure gluon sector given by \cite{anatomy}
\be \frac{\Gamma^{gl_4}}{\Gamma^{gl_3}}=\frac{\Gamma^{gl_3}}{\Gamma^{gl_2}},\ee
with
\be X_\gamma=\frac{\prod_{v\in\gamma^{[0]}}\Gamma^v}{\prod_{e\in\gamma^{[1]}_{\rm int}}\Gamma^e},\ee
where $\gamma^{[0]}$ is the set of vertices of $\gamma$ and $\gamma^{[1]}_{\rm int}$ the set of internal edges.
This simplifies in the pure gluon sector to a well-known Slavnov--Taylor identity for the renormalization of the two-, three- and four-point gluon amplitudes:
\be \frac{Z^{gl_4}}{Z^{gl_3}}=\frac{Z^{gl_3}}{Z^{gl_2}}.\label{SlTay}\ee
Here, the superscript ${}^{gl_n}$ indicates an $n$-gluon amplitude.
Note that this identity is also in accordance with unitarity and cutting symmetries, as it exhibits that the contributions
to any amputated 4-gluon amplitude renormalize consistently.

Correspondingly, we will denote by a superscript ${}^{gr_n}$ an $n$-graviton amplitude. These amplitudes span an infinite set ${\mathcal R}$.
In contrast to the situation in gauge theory, if we have any number of graviton self-couplings $gr_n$, we now need an infinite number
of renormalization constants $Z^{gr_n}$ if we are to maintain  our theory unitary.

The above Slavnov--Taylor identity was derived in \cite{anatomy} by the requirement that the sum of all graphs contributing to a given amplitude
at a fixed order of perturbation theory furnishes a generator of a sub-Hopf algebra. The same requirement
delivers now an infinite sequence of identities,
\be
\frac{\Gamma^{gr_{n+1}}}{\Gamma^{gr_{n}}}=\frac{\Gamma^{gr_{n}}}{\Gamma^{gr_{n-1}}},\label{ids}\ee
which is indeed an infinite number of identities leaving only
$\Gamma^{gr_{2}},\Gamma^{gr_{3}}$ undetermined. In the case of non-abelian gauge theory the corresponding ideal is respected by the counterterms
regarded as an element of ${\rm Spec}(G)$, which hence become an algebra map on the quotient of the Hopf algebra by this Hopf ideal.
This implies non-trivial relations between Hochschild one-cocycles beyond one loop, which iterate in the DSE to provide the desired identities for the counterterms.
In the gravity case, the situation is simpler from the viewpoint of absence of one-cocycles beyond one loop. This is compensated though by the necessity to construct elements in ${\rm Spec}(G)$ in accordance with the desired ideal.

To see this in some detail, we  identify a dressed primitive one-loop graph with $r$ external legs with an ordered
partition $\omega_r$ of an integer $r$ into integers $n_i>0$. We consider partitions up to cyclic permutations. Let
$\{\omega_r\}$ be the set of such partitions of $r$. For
$\omega\in\{\omega_r\}$ let $\|\omega\|$ be the size of the partition,
and let us consider partitions of size greater than one.
We write sometimes $r=|\omega|$.

The identification between such a partition
and a one-loop graph proceeds as follows.
A one loop graph provides say $m$ internal propagators $\phi_R([\Gamma^{gr_{2}}]^{-1})$, and $m$ internal vertices which are dressed by vertex functions $\phi_R(\Gamma^{gr_{s_i}})$, $s_i>2$.  We have $r=\sum_{i=1}^m (s_i-2)$, $n_i=s_i-2$ and identify such a graph with the partition
$\{s_1-2,s_2-2,\cdots,s_m-2\}$, as for example for $r=4=|\omega|$, $\omega=\{2,1,1\}$, $\|\omega\|=3$:
\be\Qins .\ee
We write $\gamma(\omega)$ for a one-loop graph which as a dressed graph can be identified with the partition $\omega$ and we regard partitions up to cyclic permutations.
\begin{prop}
For all $\omega,\omega^\prime$ with $|\omega|=|\omega^\prime|$, the relations \be
\prod_{j\in\omega}\Gamma^{j+2}/[\Gamma^2]^{\|\omega\|} =
\prod_{j\in\omega^\prime}\Gamma^{j+2}/[\Gamma^2]^{\|\omega^\prime\|}\ee
define a sub-Hopf algebra with Hochschild closed one-cocycles
$B_+^{1,|\omega|}=\sum_{\omega\in\{\omega_{|\omega|}\}}B_+^{\gamma(\omega)}$.
\end{prop}
Proof: By construction, the above relations define invariant charges $\phi_R(Q)$
for the $|\omega|$-point scattering amplitudes, as
\be X_\gamma=X_{\gamma^\prime}\Leftrightarrow \left(\frac{\prod_{v\in{\gamma^\prime}^{[0]}}\Gamma^v}{\prod_{e\in{\gamma^\prime}^{[1]}_{\rm int}}\Gamma^e}\right)
=\left(\frac{\prod_{v\in\gamma^{[0]}}\Gamma^v}{\prod_{e\in\gamma^{[1]}_{\rm int}}\Gamma^e}\right)\Leftrightarrow X_\gamma=\Gamma^{gr_{|\omega|}}Q^{2}
,\ee for all $\gamma=\gamma(\omega)$, $\gamma^\prime=\gamma^\prime(\omega^\prime)$ and \be Q=\frac{\Gamma^{gr_3}}{[\Gamma^{gr_2}]^{3/2}} =\frac{[\Gamma^{gr_4}]^{1/2}}{[\Gamma^{gr_2}]}=\cdots=\frac{[\Gamma^{gr_{n}}]^{1/n-2}}{[\Gamma^{gr_2}]^{n/(2n-4)}},\label{QQQ}\ee
 by construction. Hence one-cocycles $B_+^\gamma$ lead
to sub-Hopf algebras as in \cite{anatomy}.
\hfill $\Box$\\[5mm]
To determine these Hopf algebras explicitly it suffices to give their linearized coproduct
\be \Delta^\prime_{\rm lin}(c_m^{gr_n})=\sum_{j=1}^{m-1}c_j^{gr_n}\otimes c_{m-j}^{gr_n}+\sum_{j=1}^{m-1}2jc_{m-j}^Q\otimes c_j^{gr_n},\ee
where $c_m^{gr_n}$ is the sum of all $m$-loop graphs with $n$ external legs, and $c_m^Q$ is the sum of all 1PI contributions to $Q$ in any of the representations (\ref{QQQ}) above.
\subsection{KLT relations}
What we expect from the above is the ability to define a sequence of renormalization conditions
on Green functions $G^{gr_n}$, $n>3$, such that suitably defined counterterms  give an element in ${\rm Spec}(G)$ in accordance
with the above identities.

This is comparable to the situation in non-abelian gauge theory, where the renormalization condition of the four-gluon vertex is determined by the renormalization
of the gluon propagator and three-gluon vertex such that the Slavnov Taylor identity holds \cite{Celmaster}. Note though that the residue of a one-loop amplitude is independent of the choice of renormalization scheme.

If we go back to a non-abelian gauge theory, the identity (\ref{SlTay}) leads then to identities between residues of one-loop graphs which are
straightforward to check in particular in a symmetric renormalization scheme, where $s,t,u$ channels renormalize in the same way, and kinematics becomes combinatorics.

Similarly, for gravity we need an identity which connects the residues of $(n+1)$-graviton amplitudes to the $n$- and $(n-1)$-graviton amplitudes.
It suffices to consider a new external leg at zero momentum transfer as we are only interested in the residue.

At zero momentum transfer, with $V(k)_{\mu_1\mu_2;\nu_1\nu_2;\rho_1\rho_2}\equiv V(k)_{ \underline{\mu\nu\rho}}$ the tree-level vertex for a zero-momentum graviton coupling to a free propagating graviton with momentum $k$, $\Pi(k)_{\alpha_1\alpha_2;\beta_1\beta_2}\equiv\Pi(k)_{\underline{\alpha\beta}}$
the comparison of the residue in a $n+1$ and $n$ graviton one-loop amplitude reduces to a comparison
\be \Pi(k)_{\underline{\alpha\beta}}(k)\;\leftrightarrow\;\Pi(k)_{\underline{\alpha\mu}} V(k)_{\underline{\mu\nu\rho}}\Pi(k)_{\underline{\rho\beta}}.\ee

In general, this has no particular structure which allows to proceed \cite{bern}. The situation improves considerably though when one uses Feynman rules as suggested by the KLT relations \cite{bern}, which were put to use profitably in recent years relating perturbative gravity to non-abelian gauge theory \cite{bern}. For our analysis of DSE they  lead to the identity
\be \Pi(k)_{\underline{\alpha\mu}} V(k)_{\underline{\mu\nu\rho}}\Pi(k)_{\underline{\rho\beta}}
=\Pi(k)_{\underline{\alpha\beta}}\frac{k_{\nu_1}k_{\nu_2}}{k^2},\ee
using that for KLT Feynman rules one has
\bea k^2\Pi(k)_{\underline{\mu\alpha}}\Pi(k)_{\underline{\alpha\nu}} & = & \Pi_{\underline{\mu\nu}},\\
V(k)_{\underline{\mu\nu\rho}} & = & V(k)_{\mu_1\nu_1\rho_1}V(k)_{\mu_2\nu_2\rho_2}.\eea
One hence relates those residues by insertion of the scale-invariant tensor $\frac{k_{\nu_1}k_{\nu_2}}{k^2}$ into the $n$-graviton one-loop integral, which relates the residues, actually the full Mellin transforms, by contraction with a metric tensor. This immediately reduces relations between residues to combinatorics which determines the choice of a character on the Hopf algebra so that the relations (\ref{ids}) are fulfilled. The fact that there are no further primitives at higher loop order allows these relations to iterate
into the DSE. Explicit computations and an extension to ghost and matter fields is beyond the scope of this paper.

\section{Remarks and Conclusions}
We finish this short paper with a few remarks concerning the structure of theories with a power-counting as above such that propagators and vertices cancel in their contributions to the superficial degree of divergence. We  call theories with such a powercounting leg-renormalizable, to contrast them from the ordinary (loop)-renormalizable theories.
\subsection{Diffeomorphism invariance and residues}
We might accept that local geometry influences the renormalization conditions for Green functions, but the diffeomorphism invariance of residues
leaves their relation invariant in accordance with (\ref{ids}), and sits well with a the conceptual set-up of a gravity theory.
\subsection{Duality}
The loop-to-leg duality which we observe here deserves further investigation. Extending the first steps done here by incorporating matter and ghost fields, and hence combining the Hopf algebra structure above with the dual one of renormalizable fields hints towards a pairing between the loop- and leg-renormalizable theories which deserves clarification.
\subsection{Free Theory}
There is a natural resource to find other leg-renormalizable theories: start from a loop-renormalizable theory with coupling constant $g$ say and impose a non-linear field transformation. Even setting the coupling constant $g$ to zero after the transformation leaves us with an interacting theory
which indeed has vertices whose power-counting is inverse to the contributions of propagators, by construction. This should be a good starting point
to come to a better algebraic understanding of loop- and leg-renormalizable theories, and their interplay.

\end{document}